\documentclass[prl,nofootinbib,floatfix,amsmath,superscriptaddress,twocolumn]{revtex4-1}
\pdfoutput=1
\usepackage[T1]{fontenc}
\usepackage[utf8]{inputenc}
\usepackage{amssymb}
\usepackage{graphicx}
\usepackage{graphics}
\usepackage{amsmath}
\usepackage{amsthm}
\usepackage{color}
\usepackage{float}
\usepackage{import}

\def\i{{\bf i}}

\newcommand{\bea}{\begin{eqnarray}}
\newcommand{\eea}{\end{eqnarray}}
\def\bi{\begin{itemize}}
\def\ei{\end{itemize}}
\def\bc{\begin{center}}
\def\ec{\end{center}}

\def\C{\hbox{$\mit I$\kern-.7em$\mit C$}}
\def\R{\hbox{$\mit I$\kern-.6em$\mit R$}}

\def\ket#1{|#1\rangle}
\newcommand{\one}{\mbox{$1 \hspace{-1.0mm}  {\bf l}$}}

\def\ket#1{\left| #1\right>}
\def\bra#1{\left< #1\right|}

\newcommand{\proj}[1]{\ket{#1}\bra{#1}}
\newcommand{\ketbra}[2]{\ket{#1}\bra{#2}}

\usepackage{bm}
\usepackage{enumerate}
\newcommand{\expect}[1]{\left< #1 \right>}

\newcommand{\identity}{\one}

\newcommand{\trace}[2][]{\text{tr}_{#1}\left( #2 \right)}

\begin{document}
\author{M. Hebenstreit}
\affiliation{Institute for Theoretical Physics, University of
Innsbruck, Innsbruck, Austria}
\author{D. Alsina}
\affiliation{Dept. F\'\i sica Qu\`antica i Astrof\'\i sica, Universitat de Barcelona, Diagonal 645, 08028 Barcelona, Spain.}
\affiliation{Institut de Ciències del Cosmos, Universitat de Barcelona, Diagonal 645, 08028 Barcelona, Spain.\\}
\author{J. I. Latorre}
\affiliation{Dept. F\'\i sica Qu\`antica i Astrof\'\i sica, Universitat de Barcelona, Diagonal 645, 08028 Barcelona, Spain.}
\affiliation{Institut de Ciències del Cosmos, Universitat de Barcelona, Diagonal 645, 08028 Barcelona, Spain.\\}
\author{B. Kraus}
\affiliation{Institute for Theoretical Physics, University of
Innsbruck, Innsbruck, Austria}
\title{Compressed quantum computation using the IBM Quantum Experience}

\begin{abstract}

The notion of compressed quantum computation is employed to simulate
the Ising interaction of a 1D--chain consisting out of $n$ qubits using the universal IBM cloud quantum computer
running on $\log(n)$ qubits. The external field parameter that controls the quantum phase transition of this model translates into particular settings of the quantum gates that generate the circuit. We measure the magnetization, which displays the quantum phase transition, on a two--qubit system, which simulates a four--qubit Ising chain, and show its agreement with the theoretical prediction within a certain error. We also discuss the relevant point of how to assess errors when using a cloud quantum computer. As a solution, we propose to use validating circuits, that is to run independent controlled quantum circuits of similar complexity to the circuit of interest.

\end{abstract}
\maketitle

The faithful simulation of quantum systems remains one of the most interesting problems that can be addressed with a full-fledged quantum computer. Phenomena such as superconductivity in two dimensions, highly  frustrated condensed matter systems or the effect of topology in quantum systems are out of the reach of classical simulation. The emergence of new designs for quantum computation further motivates more detailed studies of mapping quantum problems to realistic quantum computation.

A particular instance of relevant physics that can be addressed with a quantum computation is the study of
quantum phase transitions \cite{sadchev}. Indeed, some systems undergo a quantum phase transition which is characterized by large quantum correlations at zero temperature. At their critical point, conformal symmetry is restored and correlations decay algebraically and become long-ranged. Furthermore, the entanglement entropy of the ground state of the system diverges in the thermodynamic limit at the phase transition. In general, such a large amount of entanglement cannot be described correctly in two dimensions by classical means.

Several experimental set--ups are currently employed to investigate quantum phase transitions \cite{lamata,bloch,monroe}. This poses the problem of designing refined experiments, which have so far tended to exploit the avenue of quantum simulation, rather than quantum computation. One of the reasons for that is that universal quantum computation, which can be used to simulate any system, is currently restricted to approximately ten qubits \cite{lanyon}. However, as we will also exploit here, certain simulations can be compressed and run on an exponentially smaller universal quantum computer.

A particularly interesting new approach to the use of quantum computers
is the advent of cloud quantum computation. The free access to run quantum circuits on a remote cloud computer opens the door to design new algorithms, to improve them by trial and error, and to refute or consolidate non-obvious ideas. It may be argued that cloud quantum computation plays a similar role to the introduction of personal computers or open mainframes in the early stages of
informatics.

At present, the availability of cloud quantum computation is limited to the IBM Quantum Experience project \cite{ibm}. It is a universal five-qubit quantum computer based on superconducting transmon qubits. The IBM quantum computer has already been tested in various ways, e.g., how well it performs in violating Mermin inequalities \cite{alsina}. Moreover, error correction codes, Fourier addition, preparing graph states, and fault tolerant circuit design have been considered \cite{devitt}.

In the present paper we test the performance of the IBM quantum computer with a compressed simulation of the transverse field 1D--Ising interaction. The quantum Ising model is an integrable model and an exact circuit can construct its ground state \cite{verstraete}. Moreover,
the notion of compressed quantum computation \cite{JoKr09} can be employed to simulate the Ising chain of $n$ qubits by using only $\log(n)$ qubits \cite{kraus}. It has also been applied to the XY-model and compressed quantum metrology \cite{BoMu13,BoKr15,BoSk16}. Moreover, the compressed simulation of the Ising spin chain (consisting of $2^5=32$ qubits) has been realized in an experiment using NMR quantum computing \cite{li}. 
On the available cloud quantum computer, it is now possible to simulate a four qubit Ising chain utilizing only two qubits. In order to realize this computation, we decompose the circuits for the compressed simulation into the available gate set. We run these circuits on the quantum computer and measure the order parameter that displays the quantum phase transition. Given that the size of the system is finite, we do observe smoothed changes of the order parameter that agree with the theoretical predictions within errors.

An important aspect to be addressed here is the assessment of errors. There are two sources of errors that have to be considered separately. First, it is necessary to run an experiment often enough so that statistical errors are reduced. This is an easy task since it only implies repetition of experiments. Second, systematic errors must be estimated. The situation here is particularly subtle, as a cloud computer is run by teams unrelated to its users. The problem of how to estimate a systematic error without knowing the detail of the computer is non-trivial. Nevertheless, an approach to the correct assessment of systematic errors can be done, using independent controlled circuits of similar complexity to the one of interest. This idea of estimating systematic errors produced by a black box might be of relevance for all future cloud quantum computation.

The outline of the remainder of the paper is the following. We first discuss some of the features and constraints of the IBM quantum computer. Then, we analyze errors occurring when single gates are applied as well as errors in more complex circuits. There, we introduce the concept of a validating circuit set in order to estimate the error of a quantum computation in case the user does not have direct access to the computer. After reviewing then the notion of compressed quantum simulation and the explicit circuits for the compressed simulation of the Ising interaction, we derive a two-qubit circuit suitable for the IBM quantum computer, which simulates a four-qubit spin chain. We present and discuss the results of the simulation and show that they agree with the theoretical prediction within the error estimated before. In the appendix, we outline how circuits simulating the eight and more-qubit spin chain can be constructed and argue that performing the computation will become possible once the announced improvements to the IBM quantum computer are implemented.

Let us begin by discussing some of the features and constraints of the IBM quantum computer, which consists currently of five qubits \cite{ibm}. The qubits are initiated in the computational basis state $\ket{0}$. As mentioned before, a limited but universal set of gates is available, namely the well known Clifford+T set. This set consists of the Pauli operators ($X$, $Y$, and $Z$), the Hadamard gate ($H$), phase gates ($S$, $S^\dagger$), $\pi/4$ gates ($T$, $T^\dagger$), as well as entangling controlled not gates ($CNOT$). Measurements in $Z$-basis as well as Bloch vector measurements (see below) are performable.
Currently, only limited classical control is available, e.g., implementing gates probabilistically is not supported. Moreover, the depth of the circuit, i.e., the number of gates that have to be applied successively and cannot be parallelized, is limited to 39. Due to the architecture of the quantum computer, one qubit, which we denote as qubit 2 in the following, plays a special role. It is the only qubit that can be target of a $CNOT$ gate. Note that the gate set is nevertheless universal. However, applying e.g. a $CNOT$ gate between qubits $i$ and $j$ (both different than $2$ here), which we denote by $CNOT(i,j)$ in the following, is very uneconomical. One first needs to swap qubit $j$ and qubit $2$ (which requires three $CNOT$s), apply $CNOT(j,2)$ and swap again qubits $j$ and $2$. As computations are naturally subject to both systematic and statistical errors, IBM provides access to a classical simulator that implements an error model of the quantum computing hardware and therefore allows simulation of a circuit before actually performing the computation.

Let us now investigate the errors, which occur in the computation. We will first analyze the errors that occur after applying a single gate and then consider those, which occur in an actual quantum computation, involving many gates. Note that the maximum allowed number of runs of one computation is limited to 8192, which allows to estimate the statistical error.

In order to get an estimation of the error that occurs after applying a single gate from the gate set, we perform the following procedure. We apply the single gate $A$ to the initial state $\rho(0)$. Ideally this would yield the state $A \ket{0}$. However, due to systematic errors, in the preparation as well as in the application of the gate, a state $\rho_A(0)$ is obtained. We perform tomography, which is of course also subject to both systematic and statistical errors, on the state $\rho_A(0)$. That is, we perform three experiments measuring $\expect{X}$, $\expect{Y}$, and $\expect{Z}$ with 8192 runs each. In order to measure $X$ and $Y$ the gates $H$ and $H S^\dagger$ are applied respectively prior the $Z$ measurement. An estimate, $\widehat{\rho_A}(0)$, of the state $\rho_A(0)$ is then determined using the direct inversion method, i.e., $\widehat{\rho_A}(0) = 1/2 \identity + 1/2 \sum_i \expect{\sigma_i} \sigma_i$, where $\{\sigma_i\}_i = \{X, Y, Z\}$. The fidelity, $F= \sqrt{\bra{0} A^\dagger  \widehat{\rho_A}(0) A \ket{0}}$, of the estimate $\widehat{\rho_A}(0)$ with respect to the ideal state $A \ket{0}$ is presented in Table \ref{tab:fidelities} for different choices of $A$.

\begin{table}[h!]
\centering
\footnotesize
  \begin{tabular}{ | c | c | c | c | c | c | c | c |}
    \hline
    Gate $A$ & $\identity$ & $H$ & $T$ & $S$ & S$^\dagger$ & $X$ & $CNOT$ \\ \hline
    Fidelity $F$ & 0.9813 & 0.9963 & 0.9961 & 0.9964 & 0.9920 & 0.9665  & 0.9794 \\ \hline
  \end{tabular}

  \caption{Fidelities of the estimate of the real state, $\widehat{\rho_A}(0)$, with respect to the ideal state $A \ket{0}$.}
  \label{tab:fidelities}
\end{table}

Note that $\widehat{\rho_A}(0)$ might not correspond to a physical state, as the length of the corresponding Bloch vector might be larger than $1$. Note further that IBM provides a Bloch measurement, which outputs a Bloch vector which is constructed in a similar way as described above. However, the Bloch vector is rescaled with the factor $1/\eta$ to take systematic errors into account. Here, $\eta$ is given by the difference of the probabilities of measuring the state $\ket{0}$ when $\ket{1}$ ($\ket{0}$) was prepared respectively, i.e., $\eta = p(0,\rho(0)) - p(0,\rho(1))$. A typical value for $1/\eta$ would be $1.05$. However IBM Bloch measurement gives results that are much more precise than those that we can produce\footnote{Note that IBM provides additional error parameters for single gates and coherence times.}.

Knowing the errors of a single gate is of course not sufficient to gain an estimate of the error obtained in an actual quantum computation, as it does not give any information about the error which accumulates during the computation due to e.g. a drift in the quantum computation. However, without knowing all the details of the experimental setup the derivation of a suitable error model is unfeasible. Due to that we propose here a different method to estimate the error, which is suitable in case the user of the quantum computer does not have direct access to it. The idea is to use a set of circuits which are approximately of the same length and complexity as the circuits of interest and whose output can be determined classically. We will call these circuits \textsl{validating circuits} in the following. They are chosen of the same length and complexity to ensure that they give rise to similar errors as the circuits of interest. Moreover, they are chosen to be classically simulatable such that the error can in fact be determined. As an example, consider a circuit of length $N$ containing $n$ $A$ gates, which are supposed to be the most erroneous ones. Then, a set of validating circuits is a set of circuits, $\{U_i\}_i$ where, for each $i$, $U_i$ contains $N$ gates in total and $n$ $A$ gates, while the other gates as well as the order in which the gates are applied may differ from the ones used in the original circuit.
Given that the outcome of these circuits can be computed classically, the error of the quantum computer running these circuits can be determined. One can then use
this error in order to estimate the error occurring in the circuit of interest, whose output cannot be computed easily.

As any computation performed on a few qubits can be simulated classically, the error can be determined directly without the use of a validating circuit set. However, once larger quantum computers become available such an approach might be very useful to estimate the expected error. Note that in order to derive the validating circuits, which have to be classically simulatable, one might use the results presented in \cite{vNest}. There, it has been shown that if two classically efficiently simulatable gate sets (strong simulation), the Clifford gates and the matchgates, are grouped in a particular way, then the output of the computation can also be simulated efficiently (weak simulation).

Here, the circuits of interest perform the compressed simulation of the Ising model, which will be derived below. Let us, for the sake of genuine error analysis, assume that the output of these circuits is unknown to us. In contrast, we assume that the output of the validating circuits is known. In order to construct them we consider two of the circuits performing the compressed simulation of the Ising model (for details see Appendix A). We keep the number of $CNOT$ and $T$ gates constant in order to keep the same complexity level, but exchange the other gates with random Clifford gates. Then, we perform a $Y$ measurement, as $Y$ is also measured in the circuit of interest, on one of the qubits. We repeat the procedure ten times obtaining 20 validating circuits in total. In Appendix A (Table \ref{tab:validation}), we present the error, $e = |\expect{Y_{measured}}  - \expect{Y_{ideal}}|$, of the 20 validating circuits. The average error is $0.122$, which is in good agreement with the experimental and theoretical results (see Fig. 2).

Let us now briefly review the notion of compressed quantum computation \cite{JoKr09}. It has been shown that matchgate circuits running on $n$-qubits, can be compressed into circuits using exponentially less qubits. Matchgates are two--qubit gates of the form $A\oplus B$, where the unitary $A$ ($B$) is acting on span$\{\ket{00},\ket{11}\}$ (span$\{\ket{01},\ket{10}\}$) respectively and the determinants of $A$ and $B$ coincide. The compression is possible, if the circuit consists of matchgates acting only on neighboring qubits, the input state is a computational basis state, and the output is the expectation value of $Z$ of a single qubit \cite{JoKr09}. It has been shown that the computational power of a $n$-qubit matchgate circuit is equivalent to that of a universal quantum computer running on only $\lceil \operatorname{log}(n) + 3\rceil$ qubits. That is, the output, which is also in the compressed computation obtained by measuring a single qubit, coincides. Moreover, the circuit size of the compressed computation coincides with the original size up to a factor $\log(n)$. An important fact to note here is that the computation is indeed performed by the quantum computer, as the allowed classical side computation is restricted to ${\cal O}(\log(n))$ space. Note that any polynomial--sized circuit that can be compressed can also be efficiently simulated classically (as a function of $n$) as the dimension of the Hilbert space corresponding to the compressed circuit is linear in $n$\footnote{Note that recently it has been shown that matchgate circuits can even be efficiently simulated classically in the case of arbitrary product states as input and arbitrary single qubit measurements on arbitrarily many output qubits, and adaptive measurements \cite{Br16}.}.

Compressed quantum simulation of the transverse field Ising model has already been realized in an experiment using NMR quantum computing \cite{li}.
Here, we also simulate this model with open boundary conditions, whose evolution is governed by the Hamiltonian
\begin{align}
	H(J) = \sum_{k=1}^{n} Z_k + J \sum_{k=1}^{n-1} X_k X_{k+1},
\end{align}
where $X_k$ ($Z_k$) denote $X$ ($Z$) acting on qubit $k$, respectively.
In the limit $n \rightarrow \infty$, the system undergoes a quantum phase transition at $J=1$ that is reflected in the discontinuity of the second derivative of the transverse magnetization.

The magnetization, $M(J)$ can be measured as follows \cite{sadchev,verstraete,kraus}. The system is initially prepared in the ground state of $H(0)$ and adiabatically evolved to the ground state of $H(J)$ by changing the parameter $J$ adiabatically. In order to perform digital adiabatic evolution over a time period $T$, the Hamiltonian $H(J)$ is discretized into $L+1$ steps. The evolution is then governed by a product of $L$ unitaries which are then approximated up to second order in $\Delta t = \frac{T}{L+1}$ using Suzuki-Trotter expansion. The evolution is indeed adiabatic and the approximation is valid if $T,\text{ }L \rightarrow \infty$ and $\Delta t \rightarrow 0$.
The transverse magnetization, $M(J)$, is obtained by measuring $Z$ on a single qubit.
As this adiabatic evolution together with the measurement of the magnetization is a matchgate circuit, the whole computation can be compressed into a universal quantum computation running on only $m=\operatorname{log}(n)$ qubits\footnote{Note that we assume here that $n$ is a power of two. Note further that due to the symmetry of the Ising model the compression to even $\operatorname{log}(n)$ qubits, instead of $\operatorname{log}(n)+3$ qubits, which are required for an arbitrary matchgate circuit, is possible.}. This exact simulation of the circuit has been shown to be as follows \cite{kraus}.
\begin{enumerate}
	\item Prepare the input state $\rho_{in} = \frac{1}{2^{m-1}} \identity^{\otimes m-1} \otimes \ket{+_y}\bra{+_y}$, where $Y \ket{+_y}=\ket{+_y}$,
	\item evolve the system up to the desired value of J by applying $W(J) = \prod_{l=1}^{L(J)} U_d R_l^T R_0^T$,
	\item measure $Y$ on qubit $m$ to obtain the magnetization $M(J) = -\trace{W(J) \rho_{in} W(J)^\dagger \; \identity \otimes Y_m}$.
\end{enumerate}
Here, the $m$--qubit unitary operators $R_0 = \identity \otimes e^{2 \Delta t Y_m}$, $R_l = [1 - \cos(\phi_l)] (\ketbra{1}{1} + \ketbra{2n}{2n}) + \cos(\phi_l) \identity + \sin(\phi_l) \sum_{k=1}^{n-1} \ketbra{2k+1}{2k} - h.c.$, and $U_d = \identity + (e^{i \phi_l} - 1)\ketbra{2n}{2n}$, where $\ket{k} = \bigotimes_{i=1}^{m} \ket{k_i}$ with $k_i$ such that $k = 1 + \sum_{i=1}^{m} k_i 2^{m-i}$, $\phi_l = 2 J_l \Delta t$, and $J_l = \frac{l}{L} J_{max}$ stem from the compression of the adiabatic evolution.

In order to perform this computation with the IBM quantum computer, we have to decompose the unitaries, which are required for the state preparation and the evolution into the Clifford+T gate set. In the following, we will outline the steps for the case of two qubits, which simulate a four-qubit spin chain. In Appendix C we explain how the computation can be performed for more qubits once some improvements of the quantum computer are available.

\begin{figure}[ht]
	\centering
	\resizebox{1.0\linewidth}{!}{\includegraphics{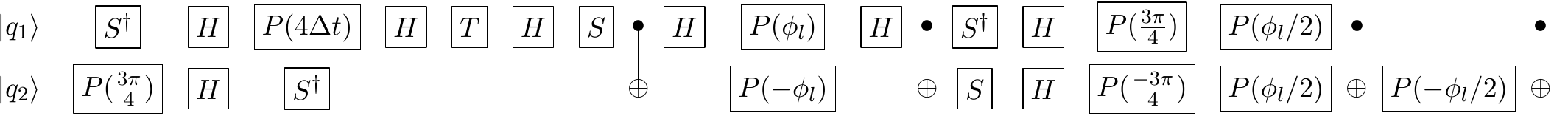}}
	\caption{Decomposition of one adiabatic step of the 2-qubit circuit into $CNOT$ and single qubit gates.}
	\label{fig:step}
\end{figure}

We exchange qubits 1 and 2 in the following due to the special role of qubit 2 in the IBM computer. The input state $\rho_{in} = \frac{1}{2} \ket{+_y}\bra{+_y} \otimes \identity$ is prepared by applying $S H$ to qubit 1 and $CNOT(3,2) H_3$ to qubit 2 and an auxiliary qubit, qubit 3, which is discarded afterwards. This procedure is uneconomical, however, it is necessary as implementing gates probabilistically is currently not possible. 
To simulate the adiabatic evolution, products of the gates $U_d, R_l^T$, and $R_0^T$ have to be applied. $R_0$ is a single qubit gate and, in the case of a two qubit circuit, $U_d=\proj{0}_1\otimes \one_2+ \proj{1}_1\otimes P_2(\phi_l)$, where $P(\phi_l)$ denotes a $\phi_l$-phase gate. The circuit depicted in Figure \ref{fig:step} implements one step in the adiabatic evolution, namely $U_d R_l R_0$, in terms of $CNOT$ and single qubit gates. Note that only the gates depending on $\phi_l$ change from step to step as $l$ is incremented in each step. The decomposition into the gate set is performed using results on decomposing arbitrary two-qubit gates into Bell diagonal gates and decomposing Bell diagonal gates into single qubit unitaries and $CNOT$ gates \cite{ViDa04, KrCi01}. All single qubit gates but phase gates depending on $\phi_l$ can be easily implemented in the Clifford+T gate set. For decomposing arbitrary phase gates we use the algorithm described in \cite{KuMa16}, where phase gates are approximated using Clifford+T gates. As there is a trade-off between the circuit depth, which is restricted here, and the quality of the approximation, we are forced to introduce a noticeable error (see Fig. 2).

The circuit depicted in Figure \ref{fig:step} has a circuit depth of 18. Hence, the total circuit, where many of these adiabatic steps have to be used before $Y_1$ is measured, exceeds the current circuit depth limit if we choose a total step number $L$ such that the evolution is indeed adiabatic. Thus, in order to keep the circuit depth feasible, we calculate the two-qubit unitary, $W(J)$, and decompose this unitary into Clifford+T gates. We approximate the single qubit unitaries as well as possible respecting the limit of the circuit depth. We provide the realized circuits in Appendix B.

In Figure \ref{fig:results} we present the results for the two-qubit circuit described above, that simulates the magnetization of a four-qubit spin chain. We measured, as in the NMR experiment \cite{li}, the magnetization for 12 values of $J$, $J=\left\{\frac{1}{6}, \frac{2}{6}, \ldots, 2 \right\}$. We also use the same parameters for the digital adiabatic evolution, $L=2400$, $\Delta t=0.1$. The solid line represents the real magnetization of the four-qubit spin chain. The black circular symbols show the theoretically obtained magnetization using digital adiabatic evolution. However, due to the restricted circuit depth, the circuit has to be approximated by a feasibly sized Clifford+T gate circuit, as described above. The dark gray, diamond shaped symbols depict the magnetization after this step, assuming that the quantum computer works perfectly. Hence, the difference between the diamond shaped and the circular symbols reflects the error made in using a feasible circuit size. Finally, the orange, filled, triangular-shaped symbols denote the actual measurement outcomes obtained using the IBM quantum computer on Sept. 9th 2016. We also provide the measurement outcomes obtained using the IBM simulator, that implements an error model of the hardware. Remarkably, there is a huge discrepancy between the output of the simulator and the actual measurement outcomes, indicating that the simulator provides pessimistic predictions here. In the figure we also illustrate the error we estimated with the validating sets. As can be seen, the results we obtain lie, on average, within the corresponding error bars. Moreover, we also reprint here the results obtained for the same simulation with a NMR quantum computer \cite{li}. There, however, a rescaling, which accounts for some of the errors has been performed. Because there the experimental data (without any rescaling) is given only for the simulation of a $2^5 = 32-$qubit spin chain, a fair comparison between these results seems to be unfeasible.
    	\begin{figure}[ht]
			\centering
			\resizebox{1.0\linewidth}{!}{\includegraphics{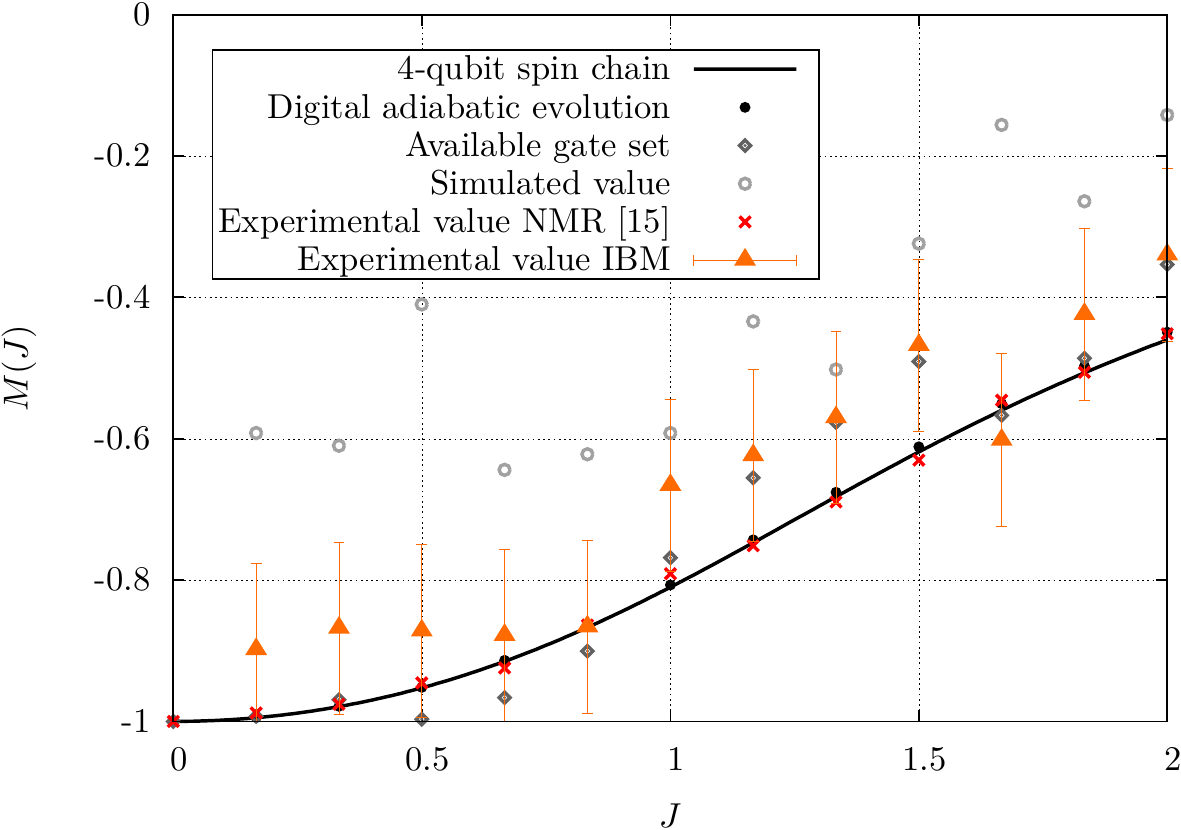}}
			\caption{The magnetization of the two-qubit circuit simulating the four-qubit spin chain (for details see main text).}
			\label{fig:results}
	\end{figure}	

In summary, we have tested the performance of the IBM quantum computer by simulating the Ising chain of four qubits using a compressed quantum simulation running on two qubits. As explained in Appendix C the realization of this simulation for more qubits will become possible once some of the announced improvements of the IBM computer will be implemented. In order to assess the error, we introduced a novel idea to estimate the error of a computation in case the user of a quantum computer does not have direct access to it. It has been shown that the obtained results agree with the theoretical predictions within this error, whereas the error estimated by the IBM simulator seems to be too pessimistic for the current experiment.

M.H. and B.K. acknowledge financial support from the Austrian Science Fund (FWF) grants Y535-N16 and DK-ALM: W1259-N27. D.A. acknowledges financial support from the APIF scholarship of University of Barcelona. J.I.L. acknowledges financial support by Grant No. FIS2013-41757-
P. We acknowledge use of the IBM Quantum Experience for this work.

The views expressed are those of the authors and do not reflect the official policy or position of IBM or the IBM Quantum Experience team.

\appendix
\section{Appendix A: Validating circuit sets}

In this section we present some details about the validating circuit sets. As explained in the main text, we introduce the concept of validating circuits in order to estimate the error that occurs in a cloud quantum computation. To this end, circuits of similar complexity as the circuit of interest, the so-called validating circuits, are considered. Assuming that the outcome of the validating circuits can be computed classically, the error is determined by comparing the real computational outcome to the ideal one. Here, we construct 20 validating circuits for the compressed simulation of the Ising model by randomly exchanging Clifford gates with other Clifford gates in circuits 2 and 3 of Figure \ref{fig:cheatingcircuits}, where the number of $T$-gates and $CNOT$-gates is not changed. We choose circuit 2 and 3 as they are of different complexity, and they together are representative for the kind of circuits that we are dealing with in simulating the Ising spin chain. 

In Table \ref{tab:validation}, we present the error $e$ of the 20 validating circuits. We perform a $Y$ measurement on one of the qubits and calculate the error given by the difference between the measured value and the ideal value, $e = |\expect{Y_{measured}}  - \expect{Y_{ideal}}|$, of the 20 validating circuits. The average error is $0.122$.

\begin{table}[h!]
\centering
\footnotesize
  \begin{tabular}{| c | c | c | c | c | c | c | c | c | c | c |}
    \hline
    $C_2$  & 0.038 & 0.076 & 0.030 & 0.130 & 0.066 & 0.166 & 0.270 & 0.128 & 0.260 & 0.000
                    \\ \hline
    $C_3$ & 0.034 & 0.202 & 0.070 & 0.152 & 0.216 & 0.076 & 0.078     &  0.248 & 0.144 & 0.056
                    \\ \hline
  \end{tabular}
 \caption{Table of the error $e$ in measuring $Y$ on one qubit in the validating circuits, which are constructed by altering two of the circuits of interest, $C_2$ and $C_3$, 10 times each.}
 \label{tab:validation}
 \end{table}

\section{Appendix B: Circuits for the simulation of the four--qubit Ising chain}
\label{app:circuits}

\begin{widetext}
 
\begin{figure}[H]
	\centering
	\resizebox{1.0\textwidth}{!}{\includegraphics{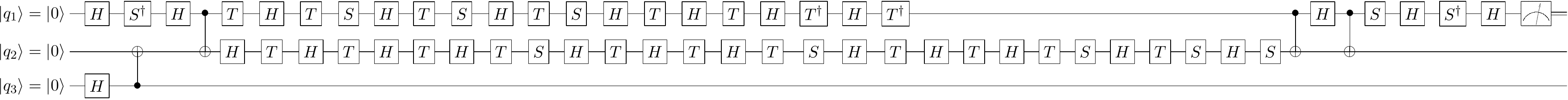}} \newline\vspace{1mm}
	\resizebox{1.0\textwidth}{!}{\includegraphics{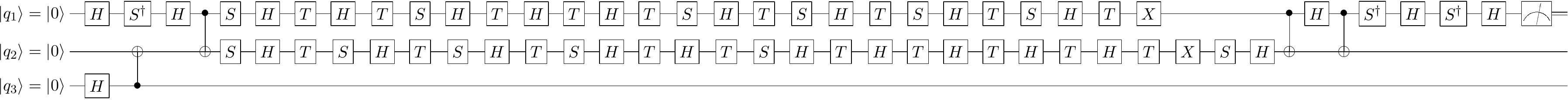}} \newline\vspace{1mm}
	\resizebox{1.0\textwidth}{!}{\includegraphics{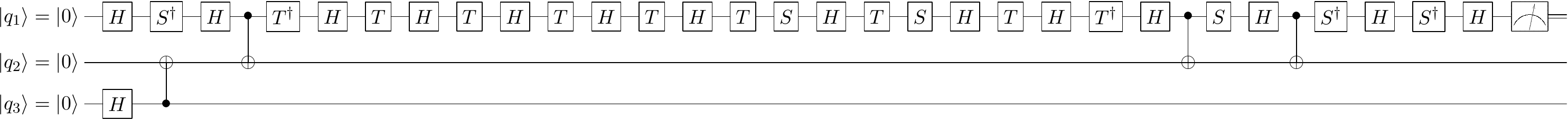}} \newline\vspace{1mm}
	\resizebox{1.0\textwidth}{!}{\includegraphics{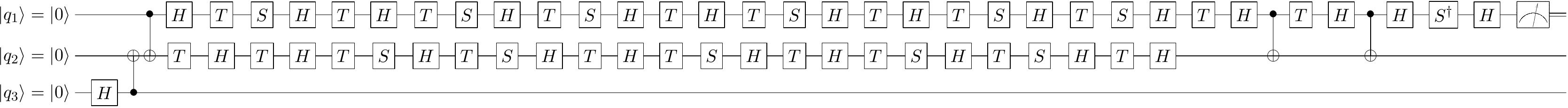}} \newline\vspace{1mm}
	\resizebox{1.0\textwidth}{!}{\includegraphics{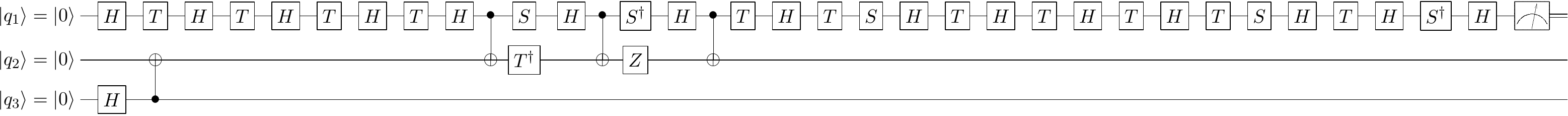}} \newline\vspace{1mm}
	\resizebox{1.0\textwidth}{!}{\includegraphics{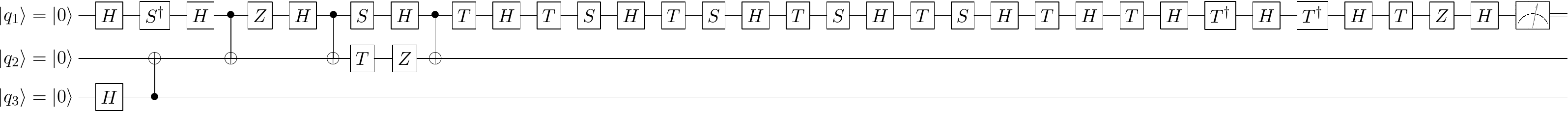}} \newline\vspace{1mm}
	\resizebox{1.0\textwidth}{!}{\includegraphics{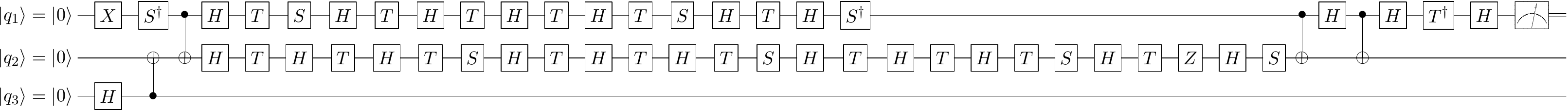}} \newline\vspace{1mm}
	\resizebox{1.0\textwidth}{!}{\includegraphics{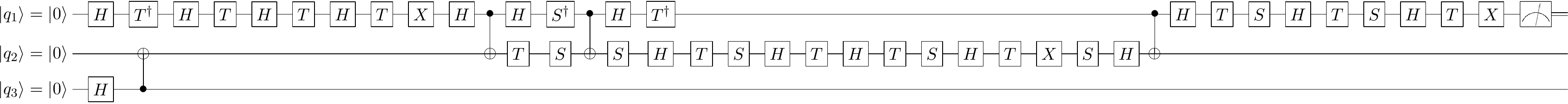}} \newline\vspace{1mm}
	\resizebox{1.0\textwidth}{!}{\includegraphics{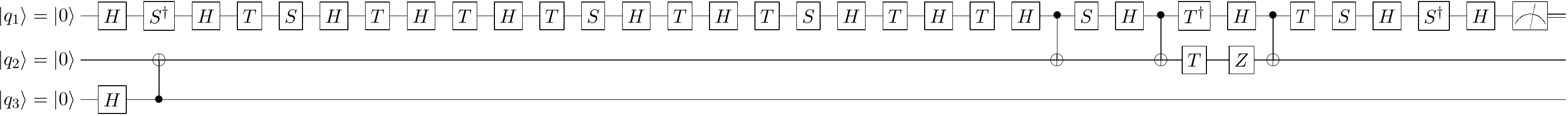}} \newline\vspace{1mm}
	\resizebox{1.0\textwidth}{!}{\includegraphics{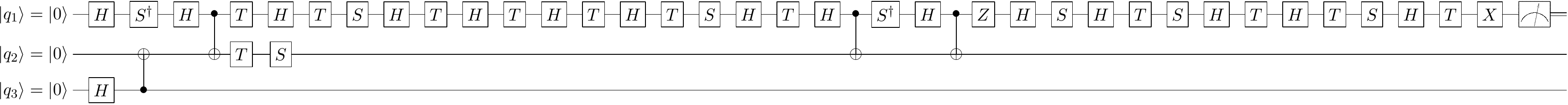}} \newline\vspace{1mm}
	\resizebox{1.0\textwidth}{!}{\includegraphics{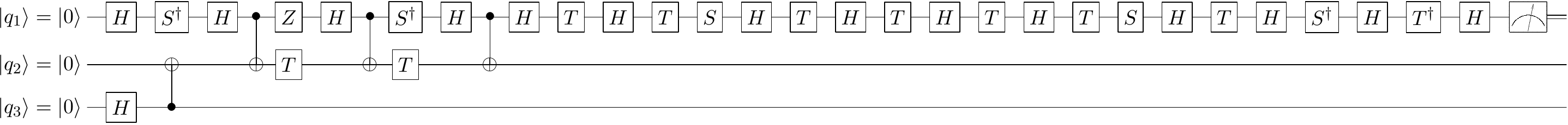}} \newline\vspace{1mm}
	\resizebox{1.0\textwidth}{!}{\includegraphics{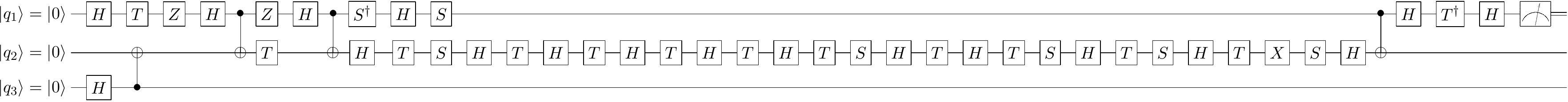}} \newline\vspace{1mm}
	\caption{Circuits implemlenting digital adiabatic evolution in order to simulate the magnetization of a four-qubit spin chain using two qubits. The twelve circuits correspond to values $J=\left\{\frac{1}{6}, \frac{2}{6}, \ldots, 2 \right\}$ as in \cite{li}. }
	\label{fig:cheatingcircuits}
\end{figure}
 \end{widetext}

In this section we explicitly give the circuits simulating the magnetization of a $4$-qubit spin chain using $2$ qubits. We measure the magnetization of the spin chain at 12 equidistantly distributed values of $J$. In particular, we choose $J=\left\{\frac{1}{6}, \frac{2}{6}, \ldots, 2 \right\}$, as in \cite{li}. We also choose the parameters of the adiabatic evolution, $\Delta t=0.1$ and $L=2400$. See main text for an explanation of these parameters. As explained in the main text, we compute the unitary $W(J)$ performing the whole adiabatic evolution and decompose this unitary into the available gates set, as a step-wise implementation of the adiabatic evolution is not possible at the moment due to the current limit in circuit depth. We entangle qubit 2 with an auxiliary qubit, qubit 3, which is discarded afterwards in order to prepare $\identity$ on qubit 2. In each circuit we measure qubit 1 in order to obtain the magnetization $M(J)$. The explicit circuits for each value $J$ are given in Figure \ref{fig:cheatingcircuits}.

\section{Appendix C: Extension to more qubits}
\label{app:extension}

In the following, we argue that with the current version of the IBM quantum computer it seems unfeasible to run the compressed simulation of the Ising spin chain using three or more qubits and hence, simulating a eight or more-qubit spin chain. Nevertheless, we show that the computation will become possible once several improvements that IBM announced are implemented.

\begin{figure}[ht]
	\centering
	\resizebox{0.6\linewidth}{!}{\includegraphics{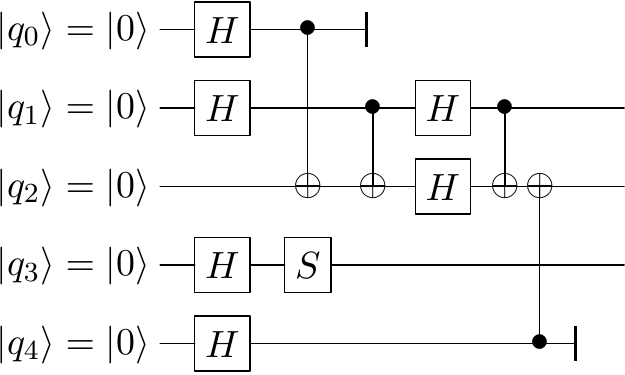}}
	\caption{Circuit for preparation of the three-qubit state $\rho_{in} = \frac{1}{4} \identity \otimes \ket{+_y}\bra{+_y}$.}
	\label{fig:preparation3}
\end{figure}

At the moment, performing the computation using three or more qubits seems not possible due to the restriction in circuit depth, the limited gate set, and the fact that gates cannot be implemented probabilistically. We exemplarily show, that even preparing the initial state $\rho_{in}$ is a difficult task. To obtain the initial state, two of the qubits have to be prepared in a completely mixed state, while one qubit is prepared in $\ket{+_y}$. See Figure \ref{fig:preparation3} for a possible, but very uneconomical way to do so using a circuit of circuit depth six and consuming two auxiliary qubits that are discarded in the process. Note that there seems to be no less wasteful way to prepare $\rho_{in}$ as applying gates probabilistically is not possible at the moment.

Nevertheless, once the improvements that IBM announced are available, implementing the circuit for more qubits will become possible. Here, we exemplarily show how to implement the circuit for three qubits, the method can be generalized to more qubits, though. To this end, we assume that the following improvements are available. We assume that advanced classical processing is available. In particular, we assume that it is possible to apply gates probabilistically. Furthermore, we assume that arbitrary single qubit gates are available and that subroutines are available, i.e., user-defined gates can be declared and used.

In this case the circuit can be implemented as follows. The initial state $\rho_{in} = \frac{1}{4} \identity \otimes \ket{+_y}\bra{+_y}$ is prepared by performing either a Pauli $X$ or $\identity$ with probability $1/2$ on both of the qubits for which we want to prepare $\frac{1}{2}\identity$ individually (which we will denote as qubits 1 and 2 in the following), and furthermore performing a single qubit unitary that rotates $\ket{0}$ to $\ket{+_y}$ for the remaining qubit, which we will denote as qubit 3 in the following.

\begin{figure}[ht]
	\centering
	\resizebox{0.6\linewidth}{!}{\includegraphics{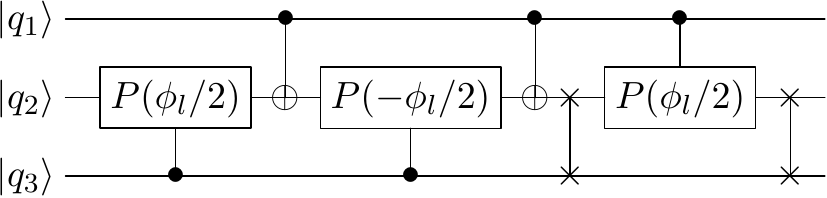}}
	\caption{Circuit implementing a $\Lambda_{1, 2} P_3(\phi_l)$ gate.}
	\label{fig:ccphase}
\end{figure}

\begin{figure}[ht]
	\centering
	\resizebox{0.8\linewidth}{!}{\includegraphics{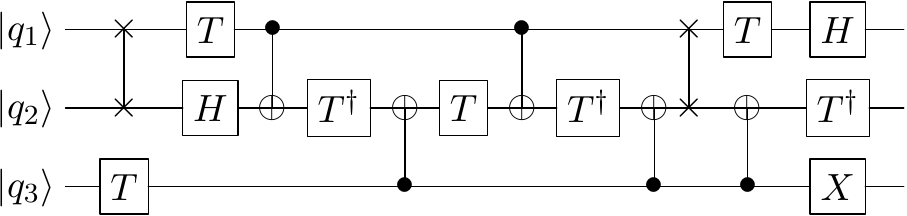}}
	\caption{Circuit implementing the operator $A$.}
	\label{fig:agate}
\end{figure}

After the initial state is prepared, the system is evolved adiabatically. In each step of this adiabatic evolution the unitary $U_d R_l^T R_0^T$ has to be applied. The unitaries $U_d$, $R_l^T$, and $R_0^T$ are given in the main text. The unitary $R_0$ is a single qubit unitary and hence can be implemented easily. We have $U_d = \Lambda_{1,2}P_3(\phi_l)$, where $\Lambda_{i_1, \ldots, i_n} G$ denotes a gate $G$ controlled by qubits $i_1, \ldots, i_n$. A possible implementation of this controlled phase gate is depicted in Figure \ref{fig:ccphase} \cite{book:nielsenchuang}. Recall that the two swaps can be implemented using three $CNOT$ gates, while phase gates that are controlled by one qubit may be decomposed into two controlled not gates and three single qubit unitaries \cite{book:nielsenchuang}.
Implementing $R_l^T$ is more tricky. First, one performs a basis transformation by applying $A^\dagger$, where $A = \ketbra{8}{1} + \sum_{k=1}^{7} \ketbra{k}{k+1}$. In the new basis the unitary $R_l^T$ is given by the unitary $\Lambda_{1,2} O^T(\phi_l)$, where $O(\phi_l) = e^{i \phi_l Y_3}$ followed by a single qubit unitary $O(\phi_l)$ \cite{BoMu13}. Finally, the basis change has to be undone, i.e., $A$ is applied. The controlled rotation can be implemented in a similar way as shown above for $U_d$. In order to implement $A$, we use that this unitary can be decomposed into a Toffoli gate $\Lambda_{2, 3} X_1$ followed by a $CNOT(2,3)$, and a Pauli $X_3$ \cite{BoSk16}. This circuit can be further decomposed using the decomposition of the Toffoli gate suitable for the IBM quantum computer \cite{ibm} and some simplifications, yielding the circuit depicted in Figure \ref{fig:agate}.

\begin{figure}[!ht]
	\centering
	\resizebox{0.8\linewidth}{!}{\includegraphics{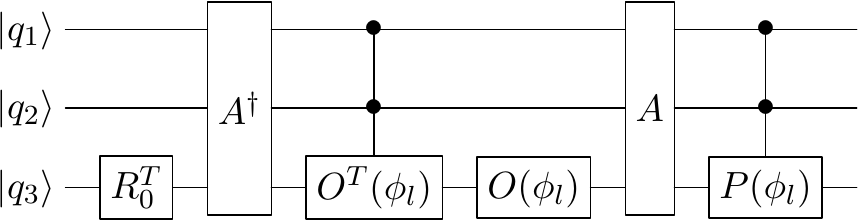}}
	\caption{Circuit implementing one step of the adiabatic evolution using three qubits.}
	\label{fig:3step}
\end{figure}

Altogether we obtain a circuit that implements one step of the adiabatic evolution depicted in Figure \ref{fig:3step}.
This circuit can be packed into a user-defined three-qubit gate depending on the free parameter $\phi_l$ and the adiabatic evolution is performed by applying these gates with increasing $l$ successively. Finally, measuring $Z$ on qubit 3 yields the magnetization of the eight-qubit spin chain.

As IBM announced, that advanced classical processing, arbitrary single qubit unitaries, and user-defined gates will become available in future, implementing the circuit for three or more qubits will become feasible, as long as the number of steps (recall that we used L=2400 steps before) is not an issue. Otherwise, similar methods as those used in the two qubit circuit will have to be applied.

\end{document}